\documentclass[12pt,preprint]{aastex}
\usepackage{natbib}
\usepackage{graphicx}
\usepackage{wrapfig}

\shorttitle{?}
\shortauthors{Araya Salvo et al.}

\begin{document}

\newcommand{\msun}{M$_{\odot}$~}
\newcommand{\etal}{{\it et al.}~}
\newcommand{\chandra}{{\it Chandra}~}
\newcommand{\chandran}{{\it Chandra}}
\newcommand{\mbh}{M$_{\rm BH}$~}
\newcommand{\xmm}{{\it XMM-Newton}~}
\newcommand{\xmmn}{{\it XMM-Newton}}
\newcommand{\mdot}{\.{m}~}
\newcommand{\mdotn}{\.{m}}
\newcommand{\Mdot}{\.{M}}
\newcommand{\es}{ergs s$^{-1}$}
\newcommand{\sig}{$\sigma$}

\title{Discovery of an active supermassive black hole in the
bulge-less galaxy NGC 4561.}

\author{C.~Araya Salvo\altaffilmark{1},
S.~Mathur\altaffilmark{1},
H.~Ghosh\altaffilmark{2},
F. Fiore\altaffilmark{3},
L.~Ferrarese\altaffilmark{4}}

\altaffiltext{1}{Department of Astronomy, The Ohio State University, 140
W 18th Ave, Columbus, OH 43210; araya@astronomy.ohio-state.edu}
\altaffiltext{2}{CNRS/CEA-Saclay, 91911 Gif-sur-Yvette, France}
\altaffiltext{3}{Osservatorio Astronomico di Roma, via Frascati 33,
100040 Monteporzio Catone, Italy} 
\altaffiltext{4}{Hertzberg Institute
of Astrophysics, 5071 West Saanich Road, Victoria, BC, V9E 2E7, Canada}

\begin{abstract}
We present \textit{XMM-Newton} observations of the \chandra-detected
nuclear X-ray source in NGC 4561. The hard X-ray spectrum can be
described by a model composed of an absorbed power-law with $\Gamma$ =
2.5$^{+0.4}_{-0.3}$, and column density $N_H=1.9^{+0.1}_{-0.2} \times
10^{22}$ atoms cm$^{-2}$. The absorption corrected luminosity of the
source is L(0.2 - 10.0 keV) $= 2.5 \times 10^{41}$ ergs s$^{-1}$, with
bolometric luminosity over $3 \times 10^{42}$ ergs s$^{-1}$. Based on
the spectrum and the luminosity, we identify the nuclear X-ray source in
NGC 4561 to be an AGN, with a black hole of mass \mbh $>
2\times10^4$ \msun. The presence of a supermassive black hole at the
center of this bulge-less galaxy shows that black hole masses are not
necessarily related to bulge properties, contrary to the general
belief. Observations such as these call into question several
theoretical models of BH--galaxy co-evolution that are based on
merger-driven BH growth; secular processes clearly play an important
role. Several emission lines are detected in the soft X-ray spectrum of
the source which can be well parametrized by an absorbed diffuse thermal
plasma with non-solar abundances of some heavy elements. Similar soft
X-ray emission is observed in spectra of Seyfert 2 galaxies and low
luminosity AGNs, suggesting an origin in the circumnuclear plasma.
\end{abstract}


\section{INTRODUCTION}
  
The past decade has seen extraordinary growth in our understanding of
supermassive black holes (SMBHs), with secure detections, mass
measurements and new demographic information (see \citealt{Ferrarese05}
and references therein). Knowledge of the mass function of SMBHs
directly affects our understanding of SMBH formation and growth, nuclear
activity, and the relation of SMBHs to the formation and evolution of
galaxies in hierarchical cold dark matter models
(e.g., \citealt{Menci04}). The cumulative mass function needed to explain
the energetics of high redshift quasars implies that all galaxies in the
local universe should host a SMBH (e.g., \citealt{Marconi04},
\citealt{Shankar04}).

Observationally, however, we do not know whether every galaxy hosts a
 SMBH. Traditional methods of finding SMBHs, viz. stellar dynamics and
 gas dynamics are powerful only at the high-mass end of the SMBH mass
 function: the BH sphere of influence cannot be resolved for BH masses
 less than $10^6$ M$_{\odot}$ beyond a distance of a couple of Mpc, even
 with \textit{HST} (\citealt{Ferrarese05}). One way, and perhaps the most
 efficient way to find SMBHs in galaxies is to look for active SMBHs. In
 fact, looking for AGN activity is perhaps the only viable way to probe
 the low-mass end of the local SMBH mass function. X-ray observations
 provide the best opportunity for this purpose because X-ray emission is
 an ubiquitous property of AGNs and X-rays can penetrate obscuring
 material that might hide an AGN at other wavelengths.  Indeed, X-ray
 observations have detected AGN activity in what were thought to be
 ``normal'' galaxies in clusters (e.g., \citealt{Martini02}) and in the
 field (e.g., \citealt{Brand05}).

In an effort to study the demographics of local SMBHs, we had undertaken
a \chandra program to look for nuclear X-ray sources in nearby optically
``normal'' spiral galaxies (within 20Mpc). This program was highly
successful; we discovered AGNs in the nuclei of what were thought to be
``normal'' galaxies (\citealt{Ghosh08}, \citealt{Ghosh09},
\citealt{Ghosh11}; see also \citealt{Zhang09};
\citealt{Desroches09}). The nuclear X-ray sources, however, could be
stars, binaries, supernova remnants or AGNs. Through extensive spectral,
timing, and multiwavelength analysis we classified the nuclear X-ray
sources and found 17 (out of 56 surveyed) that are almost certainly
low-luminosity AGNs. Thus at least 30\% of ``normal'' galaxies are
actually active. The inferred luminosities of these sources range from
$10^{37.5}$ to $10^{42}$ \es.  In a few objects where SMBH masses were
known from stellar/gas velocity dispersion methods, we find accretion
rates as low as $10^{-5}$ of the Eddington limit, comparable to what has
been found in LINERS (e.g., \citealt{Dudik05}).

We then expanded upon the initial \chandra survey by using the sample of
SINGS galaxies (Spitzer INfrared Galaxy Survey;
\citealt{Kennicutt03}). Out of the 75 SINGS galaxies, 60 have data in
the \chandra archive and we detected nuclear X-ray sources in 36 of them
(\citealt{Grier11}). This once again shows that X-ray observations are
far more efficient at detecting AGNs than optical.

  As noted above, through multiwavelength analysis we have shown that a
large fraction of the \chandra-detected nuclear X-ray sources are indeed
AGNs. Additionally, using statistical arguments we have shown most of them
to be AGNs (Ghosh et al. 2010, \citealt{Grier11}). We obtained \xmm
spectra of several of our \chandra-detected nuclear X-ray sources with
the goal of obtaining secure identifications as either AGNs or other
contaminants. In this paper we focus on the bulge-less galaxy NGC 4561
for the following reason.

The mass of the supermassive black holes (BHs) in centers of galaxies
was found to be correlated with the bulge luminosity of host galaxies
(\mbh--L$_{Bulge}$ relation; \citealt{Magorrian98}; revised in
\citealt{Gultekin09}). Even a tighter correlation was later found
between the BH mass and the velocity dispersion (\sig) of the bulge (
\mbh -- $\sigma$ relation; \citealt{Gebhardt00}, \citealt{Ferrarese00},
\citealt{Merritt01}). Basically, the mass of the black hole seems to be
correlated with the mass of the bulge (\citealt{Haring04}). The above
relations for normal galaxies also extend to active galaxies (e.g.,
\citealt{McLure02}, \citealt{Woo02}). These results were
interpreted to imply that the formation and growth of the nuclear black
hole and the bulge in a galaxy are intimately related, and several
theoretical models have attempted to explain the observed \mbh --
$\sigma$ and \mbh--L$_{Bulge}$ relations (e.g. \citealt{Adams01},
\citealt{DiMatteo03}). The hydrodynamic cosmological simulations, such
as those of \citealt{Hopkins06}, naturally account for BH--galaxy
co-evolution. In all such models, the bulge determines the nuclear BH
mass. In the models of Volonteri and Natarajan (2009) seed BH masses are
correlated to the host dark matter properties; these seeds could have
been formed by direct collapse of pre-galactic halos
(\citealt{Begelman06}).  Major mergers then trigger simultaneous BH
growth and star formation resulting in a tight coupling between the
two. A clear prediction of these models is that low-mass bulge-less
galaxies today are unlikely to host nuclear BHs (Volonteri, Natarajan \&
G\..{u}ltekin 2011). Finding AGNs in bulge-less galaxies would certainly
be a challenge for such models. Perhaps the BHs in bulge-less galaxies
represent the seed BHs that have not yet grown.  It has also been shown
that the merger and SMBH growing process may not create a bulge because
of star formation and, mainly, because of supernova feedback
(\citealt{Governato10}).  Recent studies have shown
that even moderate-luminosity AGNs up to z$\sim$3 are powered mostly through
internally driven processes (\citealt{Mullaney11}), and that mergers do
not play a major role in triggering AGNs (\citealt{Cisternas11},
\citealt{Schawinski11}). Thus the role of mergers in the growth of SMBHs
remains a matter of debate and SMBHs in bulge-less galaxies provide an
important piece of the puzzle.  Enlarging the sample of these objects is
thus crucial to learn their properties and to be able to understand
their formation and evolution mechanisms.

In this paper we present \xmm data of the nuclear X-ray source in the
bulge-less galaxy NGC 4561 which shows evidence of being an AGN. This
secure identification of a SMBH in a bulge-less galaxy shows that a bulge
is not necessary for the existence of a BH and the presence of SMBHs in
bulge-less galaxies is more common than expected.

\section{NGC 4561}

NGC 4561 (Sdm) is a late-type bulge-less spiral galaxy at z=0.00469. It
follows the selection criteria used in our previous \chandra survey
(close to face on with inclination less than $35^\circ$ to ensure that
the nuclear source is not obscured by the disk of the host galaxy;
Galactic latitude $ \arrowvert b \arrowvert > 30^\circ$ to avoid
obscuration and contamination from our own galaxy; and no known
starburst or AGN activity) in which a nuclear X-ray source was detected
(\citealt{Ghosh09}).  The optical spectra of NGC 4561 were analyzed by
\citealt{Kirhakos90}, who classified it as an H II region-like
galaxy.

The nuclear X-ray source was detected in a \chandra observation with 103
net counts with a count rate of 0.029 $\pm$ 0.003 ct/s. The source was
found to be hard ($HR = -0.53^{+0.14}_{-0.13}$).  The spectrum was
fitted with a power law $\Gamma=1.5$ $\pm$ $0.3$ and no intrinsic
absorption (N$_H \leq 1.7 \times 10^{21}$ cm$^{-2}$). With this model
the flux was F(0.3 - 8 keV) $ = 2.5 \times 10^{-13}$ ergs cm$^{-2}$
s$^{-1}$ which corresponds to a luminosity of L(0.3 - 8 keV) $\simeq 5
\times 10^{39}$ ergs s$^{-1}$. This luminosity is a lower limit: the
hardness of the source indicates that the source is likely to be
absorbed. The quality of the \chandra data, however, was not good enough
to determine the spectral shape accurately.  The hardness of the
emission and the high enough luminosity make this source a good
candidate AGN (\citealt{Ghosh09}).

The \chandra observations also showed a second source at 7'' from the
nucleus (source ``B'' here onward), which is soft ($HR=-0.9$) and its flux
is only 10\% of the nuclear source flux. Even though source B won't be
resolved by \xmm, it's effect on the spectrum of the nuclear X-ray
source should be minimal.

\section{Observations and Data Reduction}

Our target was observed with \xmm on the 2009 July 10. For
the European Photon Imaging Camera (EPIC), the exposure times were 57022
s for MOS1, 57038 s for MOS2 and 55960 s for pn; all of these exposures
were obtained using the thin filter in extended full frame.


The data were processed and filtered using SAS v9.0.0 using tasks
 \textit{epchain} and \textit{emchain}. Before the source extraction, we
 applied standard and temporal filters to the event list.  For the
 standard filter we selected the energy to be between 0.2 and 15 keV for
 pn and between 0.2 and 12 keV for MOS, single or double pixel events
 (PATTERN $<=$ 4) for pn and single, double or triple pixel events for
 MOS (PATTERN $<=$ 12), with good flag values (\#XMMEA\_EP) for pn and
 (\#XMMEA\_EM) for MOS. For the temporal filters we created light curves
 for the 10-12 keV band selecting only events with PATTERN==0 (as
 recommended at the XMM SAS User
 Guide)\footnote[1]{http://xmm.esa.int/external/xmm\_user\_support/documentation/sas\_usg/USG/}. The
 good time intervals were created by rejecting the intervals with count
 rates higher than 0.5 cts/s for pn and 0.25 cts/s for MOS. The
 effective exposure time is 26.3 ks (for pn) corresponding to 47\% of
 the observation time. The effective exposure times for MOS after
 applying the GTI are 36.7ks (64\%) for MOS1 and 37.6ks (66\%) for MOS2.

Within the circular extraction region of 20" radius the total
number of counts detected in the PN data was 6638 cts with a rate of
0.282 $\pm$ 0.004 cts s$^{-1}$. The exposure time at the source
(after vignetting correction) was 21.91 ks.  Our source was detected in
soft and hard X-rays with 4529.4 cts (0.2-2.5 keV) and 1581.7 cts
(2.5-10.0 keV), which corresponds to $HR=-0.48 \pm 0.01$. The
background level was of 2.13 cts/pixel. There were no considerable
variations on the count rate during the observation.

\subsection{The spectra}

For extracting the spectrum we used a selection area of a 20" radius
circle centered at the source and selected only events with FLAG==0 in
order to obtain a good quality spectrum.  When extracting the background
spectrum we used a square of 41" side. Finally the RMF and ARF files
were created and \textit{backscale} was ran.  Corrections for Out of
time (OoT) events or Pile up were not needed.

The spectrum was analyzed using Xspec v12.6.0 (HEASOFT). We binned the
spectra using \textit{grppha} with a minimum of 50 cts per bin for the PN 
spectrum and 40 cts per bin for the MOS1 and MOS2 spectra.

\subsubsection{Checking for possible contamination}

We checked if spectrum was contaminated. This was a possibility given
that there is a third source at 32" from the nucleus (source C here
onward), which could contaminate the source spectrum. The source
extraction radius of 26.65" encircles 90\% of the energy, so some
contamination from source C is expected. In order to see how important
this contamination was, we tried different selection areas when
extracting the spectrum: (1) a circle with a radius of 20" centered at
the source; (2) a circle with a radius of 26.65" (which encircles 90\%
of the total energy); (3) the same 26.65" circle but excluding a 26.65"
circle around source C; and (4) a 26.65" radius semicircle on the
opposite side of the source C.

\begin{figure}
\begin{center}
\includegraphics[scale=0.46]{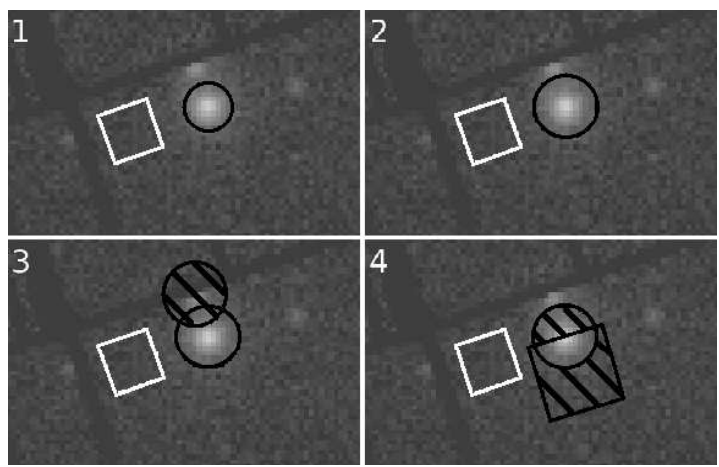}
   \caption{The EPIC-pn image of the source; the four panels show the
   different areas used to extract the spectrum when checking for
   possible contamination. To select the background spectrum the same
   41" side square (in white) was used all the time. The different
   selection areas (in black) for the spectrum are: (1) 20'' radius
   circle, (2) 26.65'' radius circle, (3) 26.65" circle radius excluding
   a circle of the same size around the closest source and (4) a 26.65''
   radius semicircle opposite to the closest source.}
\label{fig:areas}
\end{center}
 \end{figure}

These selection areas are shown in figure~\ref{fig:areas}. Emission
line-like features were present in all of the 4 extracted spectra,
discussed further in \S 4. Also, all of the spectra have similar values
for $\Gamma$ and for the observed flux showing that the contamination is
negligible, and that the emission features present in the spectrum are
real.

\subsubsection{The fitting process}

The spectral fitting was performed on the PN spectrum for energies
E$\geq$0.3 keV. Because of the considerably lower signal to noise of the
MOS data, it did not help to constrain the parameters any better. Once
the final models were chosen, we tried them on the MOS1 and MOS2
spectrum and confirmed that they are consistent.

To fit the spectra, different models were tried, always with a fixed
galactic absorption of N$_H=2.11 \times 10^{20}$ cm$^{-2}$ and a free
intrinsic absorption. Both absorption components were represented with
the ``wabs'' model for photoelectric absorption.
 
\begin{figure}
\begin{center}
\includegraphics[scale=0.42]{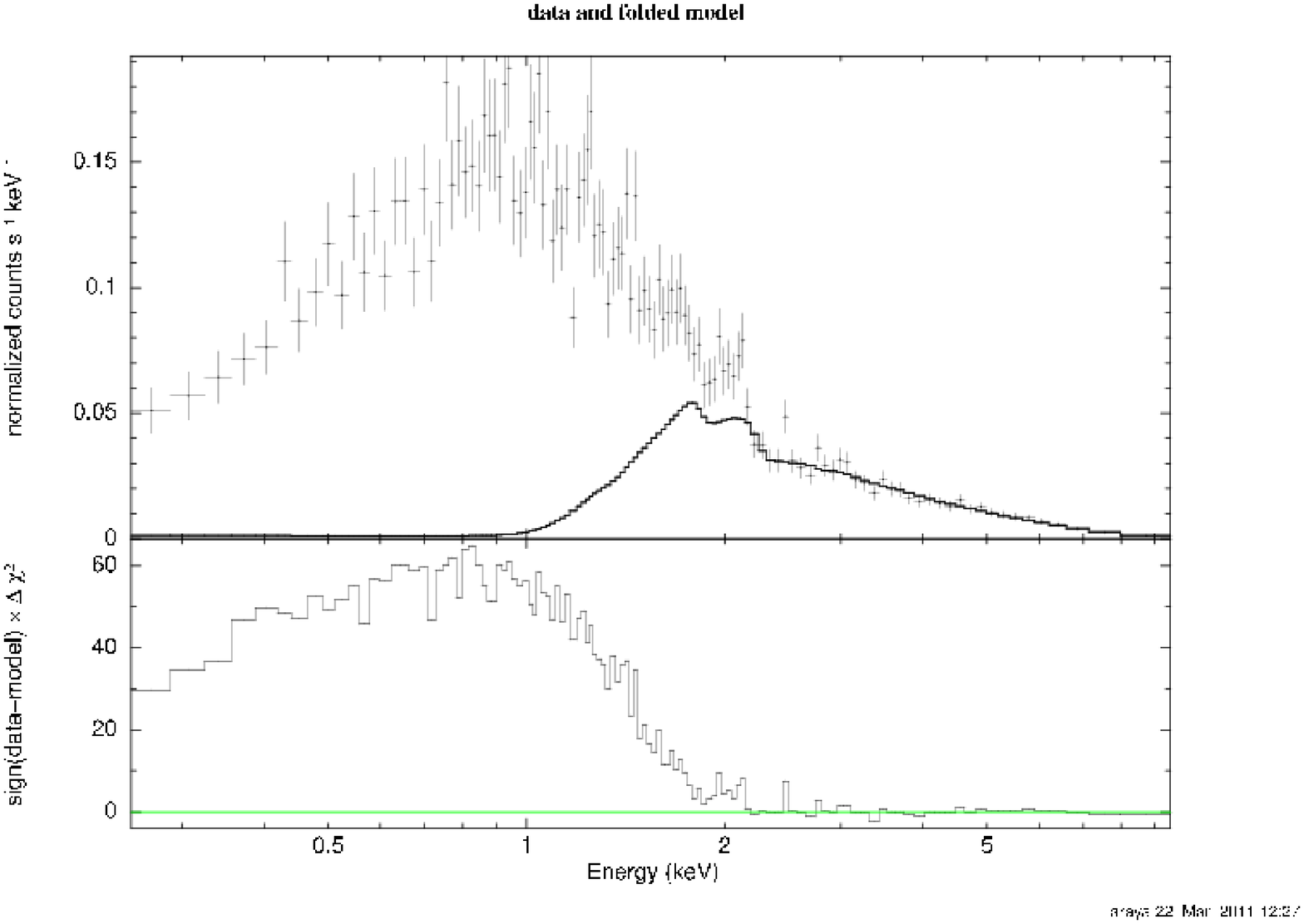}                                   
\caption{The extracted spectrum and the model of an absorbed power law
fitted to E$>$2.5 keV. The bottom panel shows the contributions to
$\chi^2$ after extrapolating the model to the entire energy
range. Significant absorption is needed to fit the hard end of the
spectrum properly.}
\label{fig:hard}
\end{center}
\end{figure}

First we tried with a power law, which by itself does not provide a good
fit ($\chi^{2}_{\nu}$=1.28 for 114 dof) and overestimates the counts for
high energies. To avoid the latter issue we fitted the absorbed power
law using only the hard part of the spectra (E$\geq$ 2.5 keV). To get a
good fit in the hard range of energies, intrinsic absorption is needed
(as suspected from the \chandra observation).  We needed another
component to model the soft-band, since the power law is almost
completely absorbed in the soft-band (see figure~\ref{fig:hard}). The
resulting photon index was $\Gamma$=2.46. From here on, when fitting the
spectrum over the entire energy range, the $\Gamma$ parameter was kept
frozen at this value.  In order to characterize the unresolved nearby
source (source B), we also added a black body component with a flux of
F(0.3-8keV)=2.5$\times$10$^{-14}$ ergs cm$^{-2}$ s$^{-1}$ corresponding
to the flux observed with \chandra and a temperature of kT=0.2 keV. Both
the temperature and the normalization were always fixed.

\begin{figure}
   \begin{center}
     \includegraphics[angle=-90,scale=0.43]{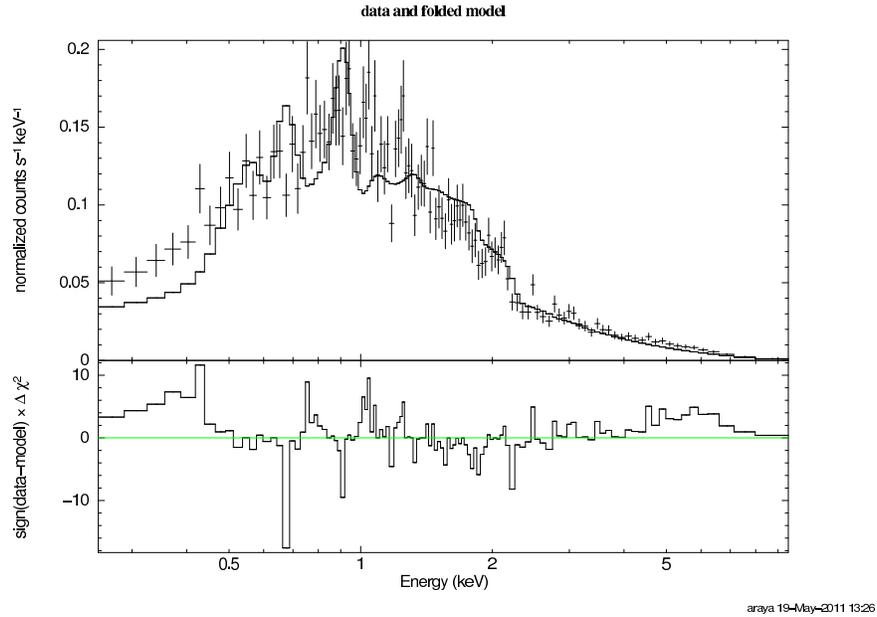}
    \end{center}
   \caption{The data fitted with an absorbed power-law plus the diffuse
   thermal plasma model (``mekal'') with solar abundances and a black
   body. Contributions to $\chi^2$ are presented in the bottom panel and
   show that this model does not fit the data well. }
   \label{fig:mekal}
 \end{figure}

\begin{figure}
   \begin{center}
     \includegraphics[angle=-90,scale=0.43]{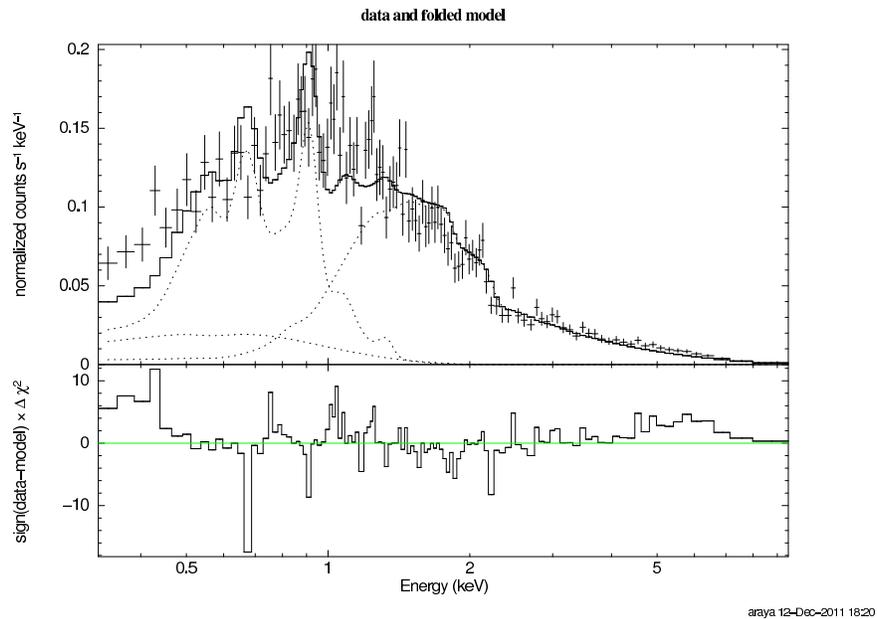}
    \end{center}
   \caption{Same as Fig. 3, but with a constant non-solar abundance
   (each component is shown with a dotted line and the solid line is the
   sum of all components).  This model does not fit the data well
   either. }
   \label{fig:mekalab}
 \end{figure}

Since the hard-band is well fit by a power-law, our next task is to fit
 the soft component with different models. Because the spectrum shows
 ``emission-like'' features, we tried an absorbed ``mekal'' model, which
 characterizes emission from hot diffuse gas, 
 but the fit was not good ($\chi^{2}_{\nu}$=2.35 for 112
 dof, even worse than the absorbed power law with $\Delta\chi^2=-117.3$,
 figure~\ref{fig:mekal}). This fit was done fixing the abundance to the
 Solar value. Leaving this parameter free does not improve the fit
 considerably ($\chi^{2}_{\nu}$=2.27 for 111 dof,
 figure~\ref{fig:mekalab}).

\begin{figure}
\begin{center}
\includegraphics[angle=-90,scale=0.44]{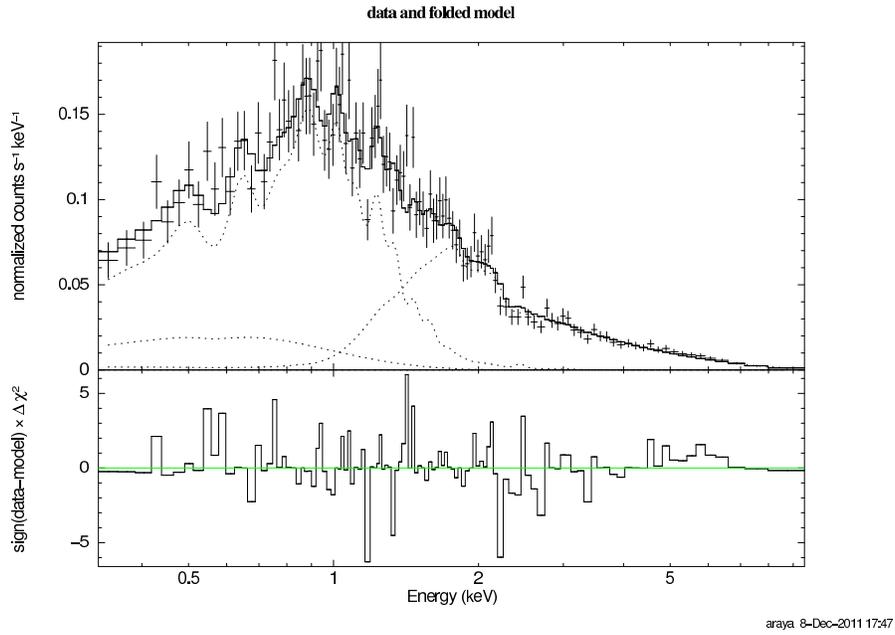}  
\caption{Same as Fig. 4, but with variable abundances. This is the
best-fit model. }
   \label{fig:thefit}
\end{center} 
 \end{figure}

We also tried to fit the soft component with the ``vmekal'' model, same
as ``mekal'' but allows the abundances of each element to vary
individually. A good fit was finally found using vmekal when leaving the
abundances of some of the heavier elements as free parameters. This
model has $\chi^{2}_{\nu}$=1.11 for 107 dof (figure~\ref{fig:thefit}),
that corresponds to $\Delta\chi^2=144.43$ when compared to the fit with
solar abundances.

\begin{figure}
   \begin{center}
     \includegraphics[angle=-90,scale=0.45]{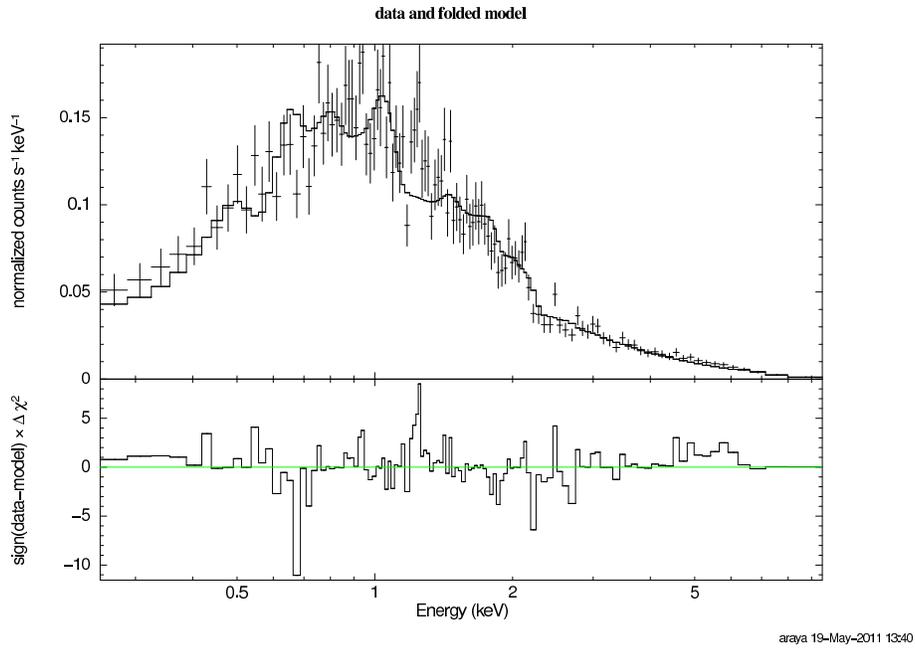}
     
    \end{center}
   \caption{The spectrum fitted with an absorbed power-law plus an
   ionized reflection component. Contributions to $\chi^2$ are presented
   in the bottom panel. This model does not provide a good fit.}
   \label{fig:reflionx}
 \end{figure}

It is possible that in the soft-band we are observing the continuum
reflected off some nearby scattering material. For this reason, we also
tried  the ``reflionx'' model (reflection by a constant density
illuminated atmosphere). This model does reproduce some emission lines, 
but the fit is not good and it is also a worse fit than the 
absorbed power law ($\chi^{2}_{\nu}$=1.39 for 113 dof,
$\Delta\chi^2=-11.18$ when compared to the absorbed power law model,
figure~\ref{fig:reflionx}).

\section{Results}

\begin{figure}
\begin{center}
\includegraphics[angle=-90,scale=0.44]{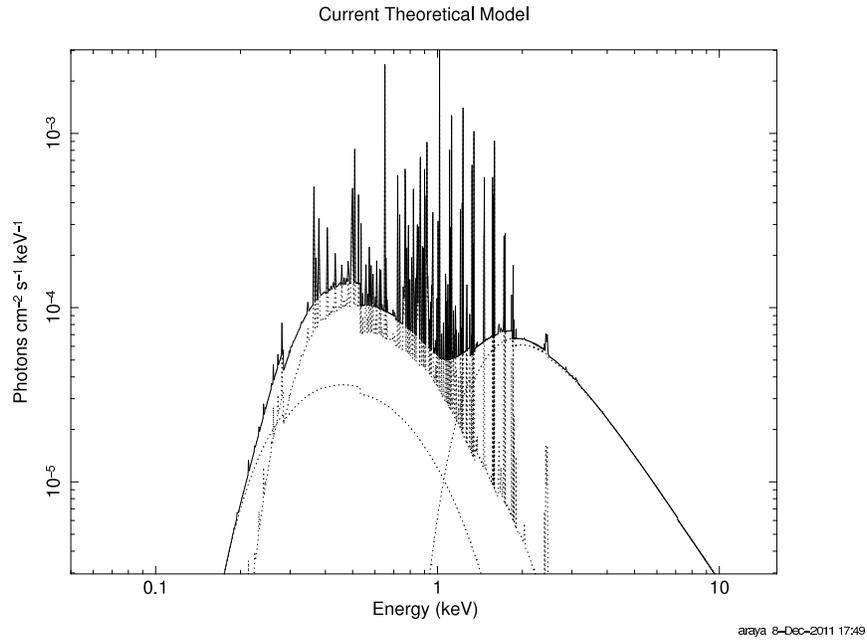}  
\caption{Theoretical model of the best fit spectrum showing the different
components (absorbed power law, absorbed thermal plasma and black
body).}
   \label{fig:uf}
\end{center} 
 \end{figure}

The best fitted model is composed of an absorbed power law with $\Gamma$
= 2.5$^{+0.4}_{-0.3}$ and $N_H=2.0^{+0.3}_{-0.2} \times10^{22}$
cm$^{-2}$ and an absorbed thermal plasma component with non-solar
abundances for some of the heavy elements (O = 0.33$^{+0.14}_{-0.12}$,
Na = 34$^{+11}_{-8}$, Si $\leq 0.2$, Fe = 0.12$^{+0.06}_{-0.04}$, Ni
$\leq0.4$) with a temperature of kT = $0.59^{+0.04}_{-0.05}$ keV and
$N_H= 7^{+2}_{-4} \times10^{20}$ cm$^{-2}$ (and a black body for the
source B). The fit to the spectrum is presented in
figure~\ref{fig:thefit} and the corresponding theoretical model in
figure~\ref{fig:uf}. The observed flux is F(0.2 - 10 keV) $ = 1.2 \times
10^{-12}$ ergs cm$^{-2}$ s$^{-1}$ which corresponds to a luminosity of
L(0.2 - 10.0 keV) $= 5.8 \times 10^{40}$ ergs s$^{-1}$. The unabsorbed
luminosity is L(0.2 - 10.0 keV) $= 2.5 \times 10^{41}$ ergs
s$^{-1}$. The bolometric luminosity of the source is therefore about
$3.5 \times 10^{42}$ ergs s$^{-1}$ (assuming a bolometric correction
factor of 14 for this luminosity; \citealt{Vasudevan07}), putting it
squarely in the luminosity range observed for AGNs.

\section{Discussion}
   
The hard-band spectrum of the nuclear X-ray source in NGC 4561 can be
described as a power law with photon index $\Gamma$ =
2.5$^{+0.4}_{-0.3}$. Most of the soft component of the power law is
absorbed (intrinsic absorption of $N_H= 2.0^{+0.3}_{-0.2}
\times10^{22}$ cm$^{-2}$). The source shows a hardness ratio of
$HR=-0.48\pm0.01$ and an absorption corrected luminosity of L(0.2 -
10.0 keV) $=2.5\times10^{41}$ ergs s$^{-1}$. The photon index and the
high luminosity of the source are indications of the presence of an
active SMBH in the center of NGC 4561.

The soft emission can be modeled as an absorbed thermal plasma with $kT=
0.59^{+0.04}_{-0.05}$ keV and non-solar abundances as follow: O =
0.33$^{+0.14}_{-0.12}$, Na = 34$^{+11}_{-8}$, Si $\leq0.2$, Fe =
0.12$^{+0.06}_{-0.04}$ and Ni $\leq0.4$, with 1.0 corresponding to the
Solar values.

\begin{figure}
\begin{center}
\includegraphics[angle=-90,scale=0.44]{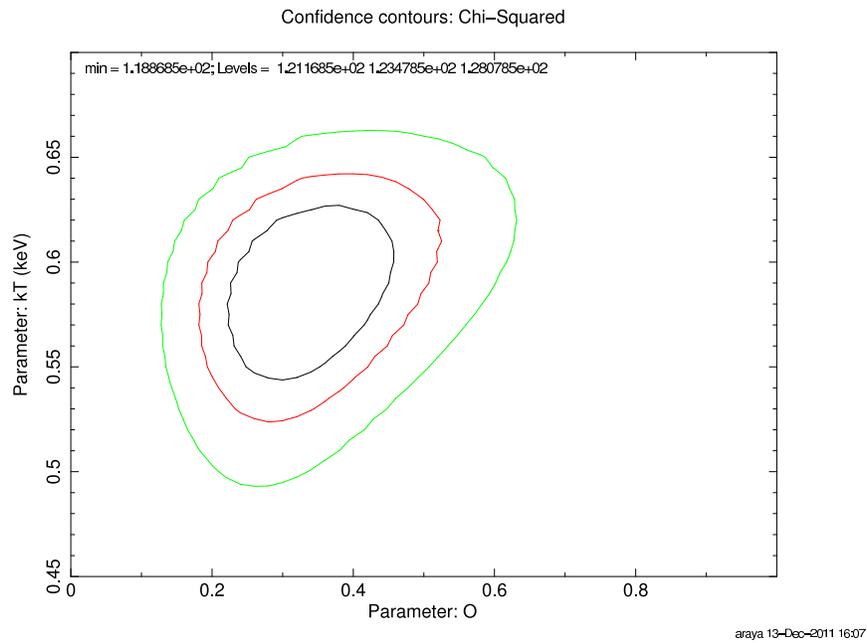}  
\caption{$\chi^2$ contours for the abundance of oxygen and temperature
of the plasma in the best fit model (Solar abundance is 1). Sub-solar
oxygen abundance is clearly required.}
   \label{fig:o}
\end{center} 
 \end{figure}

\begin{figure}
 \begin{center}
 \includegraphics[angle=-90,scale=0.44]{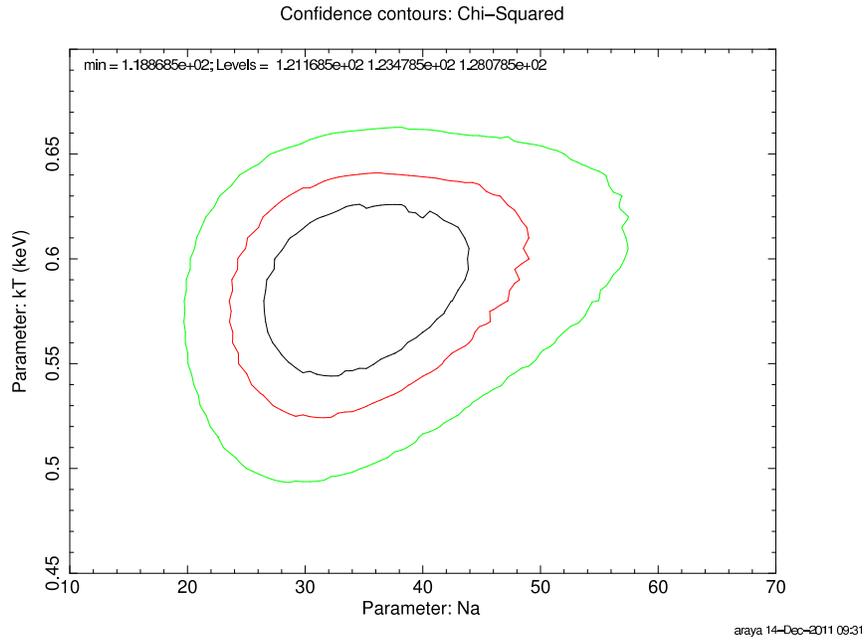}  
 \caption{$\chi^2$ contours for the abundance of sodium and temperature of the plasma in the best fit model (Solar abundance is 1). Sodium abundance appears to be super-solar.}
    \label{fig:na}
 \end{center} 
  \end{figure}

\begin{figure}
 \begin{center}
 \includegraphics[angle=-90,scale=0.44]{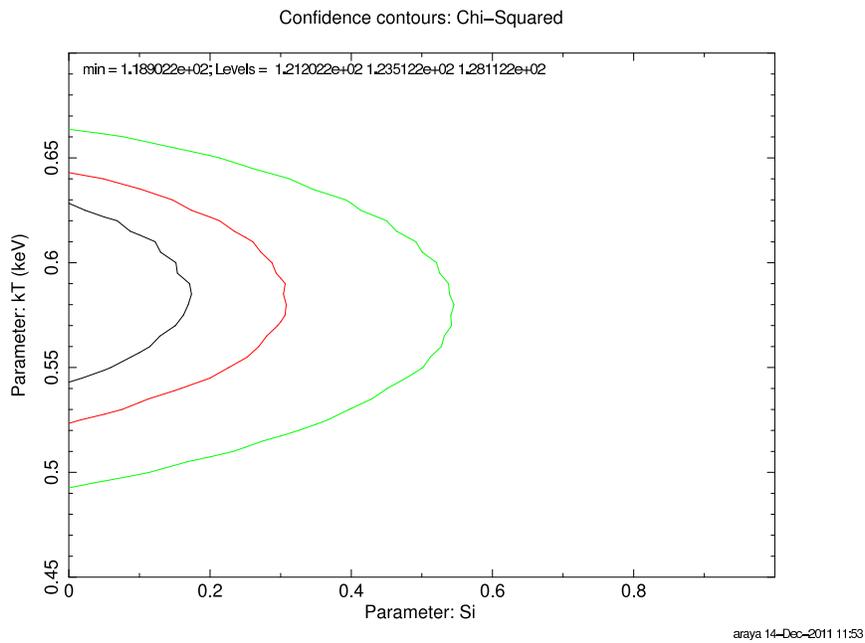}  
 \caption{$\chi^2$ contours for the abundance of silicon and temperature of the plasma in the best fit model (Solar abundance is 1). Sub-solar silicon is clearly indicated. }
    \label{fig:si}
 \end{center} 
  \end{figure}

\begin{figure}
 \begin{center}
 \includegraphics[angle=-90,scale=0.44]{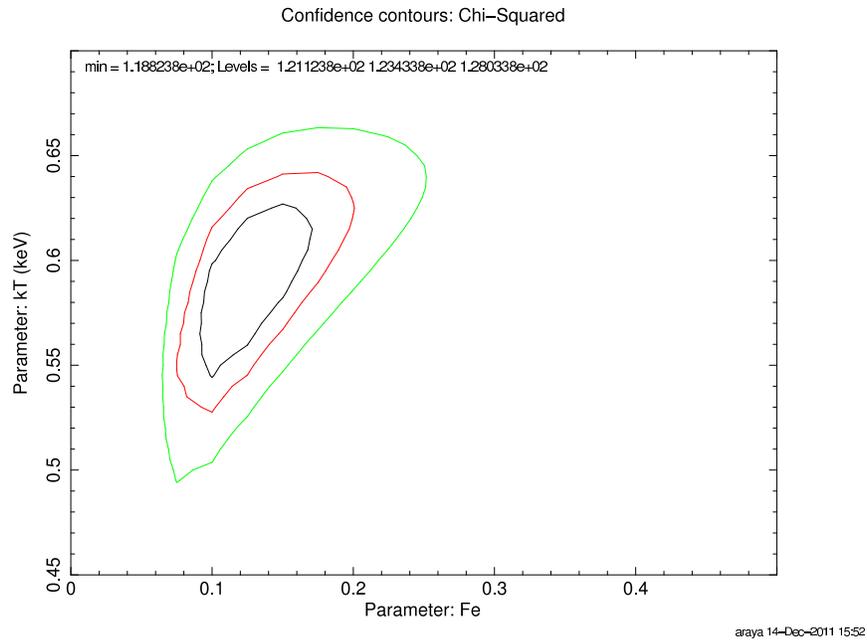}  
 \caption{$\chi^2$ contours for the abundance of iron and temperature of the plasma in the best fit model (Solar abundance is 1). Again, sub-solar iron abundance is required by the fit.}
    \label{fig:fe}
 \end{center} 
  \end{figure}

No other model was capable of providing a good fit for the spectrum. It
is quite interesting that non-solar abundances are needed for some
elements.  The nickel abundance is consistent with the solar value
within the 3\sig~ contour.  For oxygen, sodium, silicon and iron the
abundances are not consistent with the solar values as can be seen in
the contour plots (figures~\ref{fig:o}, ~\ref{fig:na},~\ref{fig:si}
and~\ref{fig:fe} for O, Na, Si and Fe respectively), they are sub-solar
for oxygen, silicon and iron but super-solar for sodium.  It is
important to note that when these parameters are kept frozen at solar
value the model does not provide a good fit. Different abundances for
different elements suggest that the observed thermal plasma is not well
mixed in heavy elements or that the models we use are too simplistic to
describe its physical conditions.

Abundances for some elements in AGNs have been measured before, for
example in Mrk 1044 and Mrk 279, where super-solar abundances for C, Ni,
O and Fe were found (\citealt{Fields05},
\citealt{Fields07}). \citealt{Hamann99} find that high-redshift quasars
also have super-solar metallicities.  Later studies showed that quasars
have super-solar abundances at all redshifts (\citealt{Hamann10}) which
would be consistent with the scenario of AGNs appearing after an
important star formation event.  Sub-solar abundances, however, have
been observed in the vicinity of AGNs like in the spectrum of NGC 1365
(\citealt{Guainazzi09}), that shows sub-solar abundances of carbon,
nitrogen, oxygen, neon, magnesium, silicon and iron. Sub-solar
abundances were inferred also for the sample of LLAGNs, LINERs and
starburst galaxies presented in \citealt{Ptak99}.

The temperature for the thermal plasma, $kT= 0.59^{+0.04}_{-0.05}$ keV,
is similar to what has been found in similar objects like NGC 3367 and
NGC 4536 ($kT= 0.64\pm0.03$ keV and $kT= 0.58\pm0.03$ keV respectively,
\citealt{McAlpine11}), and in agreement with what is expected for
LLAGNs; $kT \approx 0.4 - 0.8$ keV (\citealt{Ho08}).

\citealt{Levenson01} have analyzed ROSAT and ASCA data of Seyfert 2
galaxies with starbursts; they found that most of the soft emission,
which is modeled with a thermal component, is produced by star
formation. The median temperature of this component for their sample is
about 0.6--0.7 keV.  In their analysis, solar abundances were used
because the quality of the data would not allow a measure of
metallicity, but they do mention that low abundances are required in
high-resolution spectra of starburst galaxies as shown by
\citealt{Dahlem98}. Given the similarity of the temperature of the
thermal component and the sub-solar abundances of NGC 4561 that we find,
the soft X-ray emission we observe could be produced by nuclear star
formation.
 
No variability was observed during the observation but there was a
variation between the \chandra observation on 2006 March 15 and our \xmm
observation. From the count rate observed on \chandra (0.029 cts/s),
PIMMS predicts a count rate of 0.072 cts/s for XMM PN using the thin
filter and the whole PSF ($\sim$5 arcmin). We expect a more realistic
value of about 78\% of this rate, given the smaller (20" radius) PSF we
used; it corresponds to 0.056 cts/s. Our observed count rate of 0.282
$\pm$ 0.004 cts/s is 5 times higher than expected.  This variation might
have resulted from the change in the absorber column density. The
variability also supports the AGN scenario.

A lower limit on the black hole mass can be obtained assuming the black
hole radiates at Eddington luminosity; for L$_{\rm bol}~= 3 \times
10^{42}$ ergs s$^{-1}$, the BH must have a mass \mbh $>2\times10^4$
\msun.  This is an exciting discovery of a SMBH in a bulge-less
galaxy. To our knowledge, two bulge-less galaxies are known to host AGNs
from optical studies: NGC 4395 (Sdm; \citealt{Filippenko89},
\citealt{Peterson05}) and NGC 1042 (Scd; \citealt{Shields08}). Both
galaxies host a nuclear star cluster. IR spectroscopy with {\it Spitzer}
led to the discovery of AGNs in two more Sd galaxies, NGC 3621
(\citealt{Satyapal07}) and NGC 4178 (\citealt{Satyapal09}). These two
also harbor a nuclear star cluster. Using \xmm, 2 other AGNs in
bulge-less galaxies have been confirmed by \citealt{McAlpine11}: one in
NGC 3367 (Sc) and the other in NGC 4536 (SABbc). We found nuclear X-ray
sources in M101 (type Scd), NGC 4713 (Sd), NGC 3184 (Scd), and NGC 4561
(Sdm) (\citealt{Ghosh08}, \citealt{Ghosh09}, \citealt{Ghosh11}) and
argued that they are likely to be AGNs. The dwarf starburst galaxy
Henize 2-10 is very likely to host an AGN too (\citealt{Reines11}).  In
this paper we present conclusive evidence that NGC 4561 does in fact
host an AGN. The discovery of SMBHs in the nuclei of these galaxies
calls into question whether the masses of SMBHs are governed by bulge
properties (\S 1).  These results suggest that SMBH in bulge-less
galaxies are far more common than what we previously thought, in clear
disagreement with some models of BH growth.  A key prediction of the
models of Volonteri \& Natarajan (2009) is that low-mass bulge-less
galaxies today are unlikely to host nuclear black holes, which does not
seem to be the case.  While BHs grow through merger-driven processes,
alternative tracks of BH growth must exist; secular process appear to
play an important role.

About 75\% of late-type galaxies host nuclear star-clusters
(\citealt{Boker04}). Is the SMBH related to the mass of the star-cluster
then (\citealt{Seth08})? Or is it the dark matter halo
(\citealt{Baes03})? Are the BHs in bulge-less galaxies the seed BHs at
high redshift that did not grow?  Answering these kind of questions is
fundamental to our knowledge of BH--galaxy formation and co-evolution.

\section{Conclusions}

We present \xmm spectrum of the \chandran-detected nuclear source in NGC
4561 and show that it is an obscured AGN. The existence of nuclear SMBHs in
bulge-less galaxies shows that BH masses are not governed by
bulge properties. This calls into question several theoretical models
(\S 1) of BH--galaxy co-evolution which are merger-driven; secular
processes clearly play an important role in BH growth.

\noindent
Acknowledgment: We gratefully acknowledge support from the NASA grant
NNX09AP85G to SM.


\clearpage

\bibliographystyle{apj} 

\begin{thebibliography}{55}
\expandafter\ifx\csname natexlab\endcsname\relax\def\natexlab#1{#1}\fi

\bibitem[{{Adams} {et~al.}(2001){Adams}, {Graff}, \& {Richstone}}]{Adams01}
{Adams}, F.~C., {Graff}, D.~S., \& {Richstone}, D.~O. 2001, \apjl, 551, L31

\bibitem[{{Baes} {et~al.}(2003){Baes}, {Buyle}, {Hau}, \& {Dejonghe}}]{Baes03}
{Baes}, M., {Buyle}, P., {Hau}, G.~K.~T., \& {Dejonghe}, H. 2003, \mnras, 341,
  L44

\bibitem[{{Begelman} {et~al.}(2006){Begelman}, {Volonteri}, \&
  {Rees}}]{Begelman06}
{Begelman}, M.~C., {Volonteri}, M., \& {Rees}, M.~J. 2006, \mnras, 370, 289

\bibitem[{{B{\"o}ker} {et~al.}(2004){B{\"o}ker}, {Walcher}, {Rix},
  {H{\"a}ring}, {Schinnerer}, {Sarzi}, {van der Marel}, {Ho}, {Shields},
  {Lisenfeld}, \& {Laine}}]{Boker04}
{B{\"o}ker}, T., {Walcher}, C.~J., {Rix}, H.~W., {H{\"a}ring}, N.,
  {Schinnerer}, E., {Sarzi}, M., {van der Marel}, R.~P., {Ho}, L.~C.,
  {Shields}, J.~C., {Lisenfeld}, U., \& {Laine}, S. 2004, in Astronomical
  Society of the Pacific Conference Series, Vol. 322, The Formation and
  Evolution of Massive Young Star Clusters, ed. {H.~J.~G.~L.~M.~Lamers,
  L.~J.~Smith, \& A.~Nota}, 39--+

\bibitem[{{Brand} {et~al.}(2005){Brand}, {Dey}, {Brown}, {Watson}, {Jannuzi},
  {Najita}, {Kochanek}, {Shields}, {Fazio}, {Forman}, {Green}, {Jones},
  {Kenter}, {McNamara}, {Murray}, {Rieke}, \& {Vikhlinin}}]{Brand05}
{Brand}, K., {Dey}, A., {Brown}, M.~J.~I., {Watson}, C.~R., {Jannuzi}, B.~T.,
  {Najita}, J.~R., {Kochanek}, C.~S., {Shields}, J.~C., {Fazio}, G.~G.,
  {Forman}, W.~R., {Green}, P.~J., {Jones}, C.~J., {Kenter}, A.~T., {McNamara},
  B.~R., {Murray}, S.~S., {Rieke}, M., \& {Vikhlinin}, A. 2005, \apj, 626, 723

\bibitem[{{Cisternas} {et~al.}(2011){Cisternas}, {Jahnke}, {Inskip},
  {Kartaltepe}, {Koekemoer}, {Lisker}, {Robaina}, {Scodeggio}, {Sheth},
  {Trump}, {Andrae}, {Miyaji}, {Lusso}, {Brusa}, {Capak}, {Cappelluti},
  {Civano}, {Ilbert}, {Impey}, {Leauthaud}, {Lilly}, {Salvato}, {Scoville}, \&
  {Taniguchi}}]{Cisternas11}
{Cisternas}, M., {Jahnke}, K., {Inskip}, K.~J., {Kartaltepe}, J., {Koekemoer},
  A.~M., {Lisker}, T., {Robaina}, A.~R., {Scodeggio}, M., {Sheth}, K., {Trump},
  J.~R., {Andrae}, R., {Miyaji}, T., {Lusso}, E., {Brusa}, M., {Capak}, P.,
  {Cappelluti}, N., {Civano}, F., {Ilbert}, O., {Impey}, C.~D., {Leauthaud},
  A., {Lilly}, S.~J., {Salvato}, M., {Scoville}, N.~Z., \& {Taniguchi}, Y.
  2011, \apj, 726, 57

\bibitem[{{Dahlem} {et~al.}(1998){Dahlem}, {Weaver}, \& {Heckman}}]{Dahlem98}
{Dahlem}, M., {Weaver}, K.~A., \& {Heckman}, T.~M. 1998, \apjs, 118, 401

\bibitem[{{Desroches} \& {Ho}(2009)}]{Desroches09}
{Desroches}, L.-B., \& {Ho}, L.~C. 2009, \apj, 690, 267

\bibitem[{{Di Matteo} {et~al.}(2003){Di Matteo}, {Croft}, {Springel}, \&
  {Hernquist}}]{DiMatteo03}
{Di Matteo}, T., {Croft}, R.~A.~C., {Springel}, V., \& {Hernquist}, L. 2003,
  \apj, 593, 56

\bibitem[{{Dudik} {et~al.}(2005){Dudik}, {Satyapal}, {Gliozzi}, \&
  {Sambruna}}]{Dudik05}
{Dudik}, R.~P., {Satyapal}, S., {Gliozzi}, M., \& {Sambruna}, R.~M. 2005, \apj,
  620, 113

\bibitem[{{Ferrarese} \& {Ford}(2005)}]{Ferrarese05}
{Ferrarese}, L., \& {Ford}, H. 2005, \ssr, 116, 523

\bibitem[{{Ferrarese} \& {Merritt}(2000)}]{Ferrarese00}
{Ferrarese}, L., \& {Merritt}, D. 2000, \apjl, 539, L9

\bibitem[{{Fields} {et~al.}(2007){Fields}, {Mathur}, {Krongold}, {Williams}, \&
  {Nicastro}}]{Fields07}
{Fields}, D.~L., {Mathur}, S., {Krongold}, Y., {Williams}, R., \& {Nicastro},
  F. 2007, \apj, 666, 828

\bibitem[{{Fields} {et~al.}(2005){Fields}, {Mathur}, {Pogge}, {Nicastro},
  {Komossa}, \& {Krongold}}]{Fields05}
{Fields}, D.~L., {Mathur}, S., {Pogge}, R.~W., {Nicastro}, F., {Komossa}, S.,
  \& {Krongold}, Y. 2005, \apj, 634, 928

\bibitem[{{Filippenko} \& {Sargent}(1989)}]{Filippenko89}
{Filippenko}, A.~V., \& {Sargent}, W.~L.~W. 1989, \apjl, 342, L11

\bibitem[{{Gebhardt} {et~al.}(2000){Gebhardt}, {Bender}, {Bower}, {Dressler},
  {Faber}, {Filippenko}, {Green}, {Grillmair}, {Ho}, {Kormendy}, {Lauer},
  {Magorrian}, {Pinkney}, {Richstone}, \& {Tremaine}}]{Gebhardt00}
{Gebhardt}, K., {Bender}, R., {Bower}, G., {Dressler}, A., {Faber}, S.~M.,
  {Filippenko}, A.~V., {Green}, R., {Grillmair}, C., {Ho}, L.~C., {Kormendy},
  J., {Lauer}, T.~R., {Magorrian}, J., {Pinkney}, J., {Richstone}, D., \&
  {Tremaine}, S. 2000, \apjl, 539, L13

\bibitem[{{Ghosh}(2009)}]{Ghosh09}
{Ghosh}, H. 2009, PhD thesis, The Ohio State University

\bibitem[{{Ghosh} {et~al.}(2008){Ghosh}, {Mathur}, {Fiore}, \&
  {Ferrarese}}]{Ghosh08}
{Ghosh}, H., {Mathur}, S., {Fiore}, F., \& {Ferrarese}, L. 2008, \apj, 687, 216

\bibitem[{{Ghosh} {et~al.}(2011){Ghosh}, {Mathur}, {Fiore}, \&
  {Ferrarese}}]{Ghosh11}
---. 2011, \apj, submitted

\bibitem[{{Governato} {et~al.}(2010){Governato}, {Brook}, {Mayer}, {Brooks},
  {Rhee}, {Wadsley}, {Jonsson}, {Willman}, {Stinson}, {Quinn}, \&
  {Madau}}]{Governato10}
{Governato}, F., {Brook}, C., {Mayer}, L., {Brooks}, A., {Rhee}, G., {Wadsley},
  J., {Jonsson}, P., {Willman}, B., {Stinson}, G., {Quinn}, T., \& {Madau}, P.
  2010, \nat, 463, 203

\bibitem[{{Grier} {et~al.}(2011){Grier}, {Mathur}, {Ghosh}, \&
  {Ferrarese}}]{Grier11}
{Grier}, C.~J., {Mathur}, S., {Ghosh}, H., \& {Ferrarese}, L. 2011, \apj, 731,
  60

\bibitem[{{Guainazzi} {et~al.}(2009){Guainazzi}, {Risaliti}, {Nucita}, {Wang},
  {Bianchi}, {Soria}, \& {Zezas}}]{Guainazzi09}
{Guainazzi}, M., {Risaliti}, G., {Nucita}, A., {Wang}, J., {Bianchi}, S.,
  {Soria}, R., \& {Zezas}, A. 2009, \aap, 505, 589

\bibitem[{{G{\"u}ltekin} {et~al.}(2009){G{\"u}ltekin}, {Richstone}, {Gebhardt},
  {Lauer}, {Tremaine}, {Aller}, {Bender}, {Dressler}, {Faber}, {Filippenko},
  {Green}, {Ho}, {Kormendy}, {Magorrian}, {Pinkney}, \& {Siopis}}]{Gultekin09}
{G{\"u}ltekin}, K., {Richstone}, D.~O., {Gebhardt}, K., {Lauer}, T.~R.,
  {Tremaine}, S., {Aller}, M.~C., {Bender}, R., {Dressler}, A., {Faber}, S.~M.,
  {Filippenko}, A.~V., {Green}, R., {Ho}, L.~C., {Kormendy}, J., {Magorrian},
  J., {Pinkney}, J., \& {Siopis}, C. 2009, \apj, 698, 198

\bibitem[{{Hamann} \& {Ferland}(1999)}]{Hamann99}
{Hamann}, F., \& {Ferland}, G. 1999, \araa, 37, 487

\bibitem[{{Hamann} \& {Simon}(2010)}]{Hamann10}
{Hamann}, F., \& {Simon}, L.~E. 2010, in IAU Symposium, Vol. 265, IAU
  Symposium, ed. {K.~Cunha, M.~Spite, \& B.~Barbuy}, 171--178

\bibitem[{{H{\"a}ring} \& {Rix}(2004)}]{Haring04}
{H{\"a}ring}, N., \& {Rix}, H.-W. 2004, \apjl, 604, L89

\bibitem[{{Ho}(2008)}]{Ho08}
{Ho}, L.~C. 2008, \araa, 46, 475

\bibitem[{{Hopkins} {et~al.}(2006){Hopkins}, {Hernquist}, {Cox}, {Di Matteo},
  {Robertson}, \& {Springel}}]{Hopkins06}
{Hopkins}, P.~F., {Hernquist}, L., {Cox}, T.~J., {Di Matteo}, T., {Robertson},
  B., \& {Springel}, V. 2006, \apjs, 163, 1

\bibitem[{{Kennicutt} {et~al.}(2003){Kennicutt}, {Armus}, {Bendo}, {Calzetti},
  {Dale}, {Draine}, {Engelbracht}, {Gordon}, {Grauer}, {Helou}, {Hollenbach},
  {Jarrett}, {Kewley}, {Leitherer}, {Li}, {Malhotra}, {Regan}, {Rieke},
  {Rieke}, {Roussel}, {Smith}, {Thornley}, \& {Walter}}]{Kennicutt03}
{Kennicutt}, Jr., R.~C., {Armus}, L., {Bendo}, G., {Calzetti}, D., {Dale},
  D.~A., {Draine}, B.~T., {Engelbracht}, C.~W., {Gordon}, K.~D., {Grauer},
  A.~D., {Helou}, G., {Hollenbach}, D.~J., {Jarrett}, T.~H., {Kewley}, L.~J.,
  {Leitherer}, C., {Li}, A., {Malhotra}, S., {Regan}, M.~W., {Rieke}, G.~H.,
  {Rieke}, M.~J., {Roussel}, H., {Smith}, J.-D.~T., {Thornley}, M.~D., \&
  {Walter}, F. 2003, \pasp, 115, 928

\bibitem[{{Kirhakos} \& {Steiner}(1990)}]{Kirhakos90}
{Kirhakos}, S.~D., \& {Steiner}, J.~E. 1990, \aj, 99, 1722

\bibitem[{{Koopmann} {et~al.}(2006){Koopmann}, {Haynes}, \&
  {Catinella}}]{Koopmann06}
{Koopmann}, R.~A., {Haynes}, M.~P., \& {Catinella}, B. 2006, \aj, 131, 716

\bibitem[{{Kormendy} \& {Bender}(2011)}]{Kormendy11}
{Kormendy}, J., \& {Bender}, R. 2011, \nat, 469, 377

\bibitem[{{Levenson} {et~al.}(2001){Levenson}, {Weaver}, \&
  {Heckman}}]{Levenson01}
{Levenson}, N.~A., {Weaver}, K.~A., \& {Heckman}, T.~M. 2001, \apj, 550, 230

\bibitem[{{Magorrian} {et~al.}(1998){Magorrian}, {Tremaine}, {Richstone},
  {Bender}, {Bower}, {Dressler}, {Faber}, {Gebhardt}, {Green}, {Grillmair},
  {Kormendy}, \& {Lauer}}]{Magorrian98}
{Magorrian}, J., {Tremaine}, S., {Richstone}, D., {Bender}, R., {Bower}, G.,
  {Dressler}, A., {Faber}, S.~M., {Gebhardt}, K., {Green}, R., {Grillmair}, C.,
  {Kormendy}, J., \& {Lauer}, T. 1998, \aj, 115, 2285

\bibitem[{{Marconi} {et~al.}(2004){Marconi}, {Risaliti}, {Gilli}, {Hunt},
  {Maiolino}, \& {Salvati}}]{Marconi04}
{Marconi}, A., {Risaliti}, G., {Gilli}, R., {Hunt}, L.~K., {Maiolino}, R., \&
  {Salvati}, M. 2004, \mnras, 351, 169

\bibitem[{{Martini} {et~al.}(2002){Martini}, {Kelson}, {Mulchaey}, \&
  {Trager}}]{Martini02}
{Martini}, P., {Kelson}, D.~D., {Mulchaey}, J.~S., \& {Trager}, S.~C. 2002,
  \apjl, 576, L109

\bibitem[{{McAlpine} {et~al.}(2011){McAlpine}, {Satyapal}, {Gliozzi}, {Cheung},
  {Sambruna}, \& {Eracleous}}]{McAlpine11}
{McAlpine}, W., {Satyapal}, S., {Gliozzi}, M., {Cheung}, C.~C., {Sambruna},
  R.~M., \& {Eracleous}, M. 2011, \apj, 728, 25

\bibitem[{{McLure} \& {Dunlop}(2002)}]{McLure02}
{McLure}, R.~J., \& {Dunlop}, J.~S. 2002, \mnras, 331, 795

\bibitem[{{Menci} {et~al.}(2004){Menci}, {Fiore}, {Perola}, \&
  {Cavaliere}}]{Menci04}
{Menci}, N., {Fiore}, F., {Perola}, G.~C., \& {Cavaliere}, A. 2004, \apj, 606,
  58

\bibitem[{{Merritt} \& {Ferrarese}(2001)}]{Merritt01}
{Merritt}, D., \& {Ferrarese}, L. 2001, \apj, 547, 140

\bibitem[{{Mullaney} {et~al.}(2011){Mullaney}, {Pannella}, {Daddi},
  {Alexander}, {Elbaz}, {Hickox}, {Bournaud}, {Altieri}, {Aussel}, {Coia},
  {Dannerbauer}, {Dasyra}, {Dickinson}, {Hwang}, {Kartaltepe}, {Leiton},
  {Magdis}, {Magnelli}, {Popesso}, {Valtchanov}, {Bauer}, {Brandt}, {Del Moro},
  {Hanish}, {Ivison}, {Juneau}, {Luo}, {Lutz}, {Sargent}, {Scott}, \&
  {Xue}}]{Mullaney11}
{Mullaney}, J.~R., {Pannella}, M., {Daddi}, E., {Alexander}, D.~M., {Elbaz},
  D., {Hickox}, R.~C., {Bournaud}, F., {Altieri}, B., {Aussel}, H., {Coia}, D.,
  {Dannerbauer}, H., {Dasyra}, K., {Dickinson}, M., {Hwang}, H.~S.,
  {Kartaltepe}, J., {Leiton}, R., {Magdis}, G., {Magnelli}, B., {Popesso}, P.,
  {Valtchanov}, I., {Bauer}, F.~E., {Brandt}, W.~N., {Del Moro}, A., {Hanish},
  D.~J., {Ivison}, R.~J., {Juneau}, S., {Luo}, B., {Lutz}, D., {Sargent},
  M.~T., {Scott}, D., \& {Xue}, Y.~Q. 2011, \mnras, 1756

\bibitem[{{Paturel} {et~al.}(2003){Paturel}, {Theureau}, {Bottinelli},
  {Gouguenheim}, {Coudreau-Durand}, {Hallet}, \& {Petit}}]{Paturel03}
{Paturel}, G., {Theureau}, G., {Bottinelli}, L., {Gouguenheim}, L.,
  {Coudreau-Durand}, N., {Hallet}, N., \& {Petit}, C. 2003, \aap, 412, 57

\bibitem[{{Peterson} {et~al.}(2005){Peterson}, {Bentz}, {Desroches},
  {Filippenko}, {Ho}, {Kaspi}, {Laor}, {Maoz}, {Moran}, {Pogge}, \&
  {Quillen}}]{Peterson05}
{Peterson}, B.~M., {Bentz}, M.~C., {Desroches}, L.-B., {Filippenko}, A.~V.,
  {Ho}, L.~C., {Kaspi}, S., {Laor}, A., {Maoz}, D., {Moran}, E.~C., {Pogge},
  R.~W., \& {Quillen}, A.~C. 2005, \apj, 632, 799

\bibitem[{{Ptak} {et~al.}(1999){Ptak}, {Serlemitsos}, {Yaqoob}, \&
  {Mushotzky}}]{Ptak99}
{Ptak}, A., {Serlemitsos}, P., {Yaqoob}, T., \& {Mushotzky}, R. 1999, \apjs,
  120, 179

\bibitem[{{Reines} {et~al.}(2011){Reines}, {Sivakoff}, {Johnson}, \&
  {Brogan}}]{Reines11}
{Reines}, A.~E., {Sivakoff}, G.~R., {Johnson}, K.~E., \& {Brogan}, C.~L. 2011,
  \nat, 470, 66

\bibitem[{{Satyapal} {et~al.}(2009){Satyapal}, {B{\"o}ker}, {Mcalpine},
  {Gliozzi}, {Abel}, \& {Heckman}}]{Satyapal09}
{Satyapal}, S., {B{\"o}ker}, T., {Mcalpine}, W., {Gliozzi}, M., {Abel}, N.~P.,
  \& {Heckman}, T. 2009, \apj, 704, 439

\bibitem[{{Satyapal} {et~al.}(2007){Satyapal}, {Vega}, {Heckman}, {O'Halloran},
  \& {Dudik}}]{Satyapal07}
{Satyapal}, S., {Vega}, D., {Heckman}, T., {O'Halloran}, B., \& {Dudik}, R.
  2007, \apjl, 663, L9

\bibitem[{{Schawinski} {et~al.}(2011){Schawinski}, {Treister}, {Urry},
  {Cardamone}, {Simmons}, \& {Yi}}]{Schawinski11}
{Schawinski}, K., {Treister}, E., {Urry}, C.~M., {Cardamone}, C.~N., {Simmons},
  B., \& {Yi}, S.~K. 2011, \apjl, 727, L31

\bibitem[{{Seth} {et~al.}(2008){Seth}, {Ag{\"u}eros}, {Lee}, \&
  {Basu-Zych}}]{Seth08}
{Seth}, A., {Ag{\"u}eros}, M., {Lee}, D., \& {Basu-Zych}, A. 2008, \apj, 678,
  116

\bibitem[{{Shankar} {et~al.}(2004){Shankar}, {Salucci}, {Granato}, {De Zotti},
  \& {Danese}}]{Shankar04}
{Shankar}, F., {Salucci}, P., {Granato}, G.~L., {De Zotti}, G., \& {Danese}, L.
  2004, \mnras, 354, 1020

\bibitem[{{Shields} {et~al.}(2008){Shields}, {Walcher}, {B{\"o}ker}, {Ho},
  {Rix}, \& {van der Marel}}]{Shields08}
{Shields}, J.~C., {Walcher}, C.~J., {B{\"o}ker}, T., {Ho}, L.~C., {Rix}, H.-W.,
  \& {van der Marel}, R.~P. 2008, \apj, 682, 104

\bibitem[{{Vasudevan} \& {Fabian}(2007)}]{Vasudevan07}
{Vasudevan}, R.~V., \& {Fabian}, A.~C. 2007, \mnras, 381, 1235

\bibitem[{{Volonteri} {et~al.}(2009){Volonteri} \& {Natarajan}}]{Volonteri09}
{Volonteri}, M., \& {Natarajan}, P., 2009, MNRAS, 400, 1911

\bibitem[{{Volonteri} {et~al.}(2011){Volonteri}, {Natarajan}, \&
  {G{\"u}ltekin}}]{Volonteri11}
{Volonteri}, M., {Natarajan}, P., \& {G{\"u}ltekin}, K. 2011, \apj, 737, 50

\bibitem[{{Woo} \& {Urry}(2002)}]{Woo02}
{Woo}, J.-H., \& {Urry}, C.~M. 2002, \apj, 579, 530

\bibitem[{{Zhang} {et~al.}(2009){Zhang}, {Soria}, {Zhang}, {Swartz}, \&
  {Liu}}]{Zhang09}
{Zhang}, W.~M., {Soria}, R., {Zhang}, S.~N., {Swartz}, D.~A., \& {Liu}, J.~F.
  2009, \apj, 699, 281

\end{thebibliography}

\clearpage
   
    \begin{deluxetable}{clcccc} 
\tabletypesize{\scriptsize}
 \setlength{\tabcolsep}{0.04in}
 \tablecaption{Spectral Models} 
 \tablewidth{0pt} 
 \tablehead{ 
  \colhead{Model ID} &
 \colhead{Model} & 
\colhead{Component} &
 \colhead{Parameters\tablenotemark{*}\tablenotemark{**}}   &
 \colhead{$\chi^{2}_{\nu}$ (dof) } &  \\   
 } 
 \startdata

1 & absorbed power-law and thermal	& power-law	& N$_{H}= 1.02\pm0.05$					& 2.35 (112)  	\\
  & plasma with solar abundance	& mekal		& N$_{H}= 0.86\pm0.05$, kT $=0.11\pm0.01$				&		\\
  &					&		&							&		\\
2 & absorbed power-law and thermal	& power-law	& N$_{H}=1.06\pm0.05$ 					&  2.27 (111)  	\\
  & plasma with non-solar abundance	& mekal		& N$_{H}=0.82\pm0.05$ , $kT=0.12\pm0.01$, abund$=0.3\pm0.3$				&		\\
  &					&		&							&		\\
3 & absorbed power-law and  		& power-law 	& N$_{H}=1.6\pm0.2$ 						& 1.39 (113) 	\\
  & reflection				& reflionx 	& N$_{H}= 0.23^{+0.03}_{-0.02}$, Fe/solar = 4.5$^{+1.3}_{-2.5}$, X$_{i}=1300^{+700}_{-600}$	&		\\
  &					&		&							&		\\
4 & absorbed power-law  		& power-law	& $N_{H}=1.39\pm0.06$					& 2.35 (112)	\\
  & and vmekal with solar abundance 	& vmekal	& $N_{H}=0.0^{+0.03}$, $kT= 0.70\pm0.01$	& 		\\
  &					&		&							&		\\
5 & absorbed power-law and   		& power-law	& $N_{H}= 1.9^{+0.1}_{-0.2}$				& 1.11 (107)	\\
  & vmekal with non-solar abundance		& vmekal	& $N_{H}= 0.07^{+0.02}_{-0.03}$, $kT= 0.59^{+0.04}_{-0.05}$	&	\\
  &					&		& O = 0.33$^{+0.14}_{-0.12}$, Na = 34$^{+11}_{-8}$  & \\
  &					&		& Si = 0$^{+0.2}_{-0}$, Fe = 0.12$^{+0.06}_{-0.04}$, Ni = 0$^{+0.4}_{-0}$ & \\

 \enddata
 \tablenotetext{*}{In all the models: Galactic absorption was added with N$_{H}$= 2.11 $\times$ 10$^{20}$ cm$^{-2}$ and the photon index of the power-law is $\Gamma$ = 2.46 (obtained fitting only for E $\geq$ 2.5 keV). These two parameters were always frozen when fitting. All the models also have a blackbody with kT=0.2 keV and flux F(0.2-10 keV)=2.5$\times10^{-14}$ accounting for the source B.}
\tablenotetext{**}{Al column densities are in units of 10$^{22}$ cm$^{-2}$, temperatures in keV. }
 \label{Table:models}
 \end{deluxetable}

%
%

\end{document}